\documentclass[aps,prb,twocolumn]{revtex4}
\usepackage{tabularx}
\usepackage{graphicx}
\usepackage{amsmath,amsthm,amssymb}

\DeclareMathOperator{\sign}{sign}
\providecommand{\vect}[1]{\boldsymbol{#1}}

\begin{document}

\title{%
Topological insulators in magnetic fields:
Quantum Hall effect and edge channels with non-quantized $\theta$-term
}

\author{M.~Sitte$^1$}
\author{A.~Rosch$^1$}
\author{E. Altman$^2$}
\author{L.~Fritz$^1$}

\affiliation{%
$^1$ Institute for Theoretical Physics, University of Cologne,  Cologne, Germany\\
$^2$ Department of Condensed Matter Physics, The Weizmann Institute of Science, 76100 Rehovot, Israel
}

\date{\today}

\begin{abstract}
We investigate how a magnetic field induces one-dimensional edge channels when the two-dimensional surface states of three-dimensional topological insulators become gapped.
The Hall effect, measured by contacting those channels, remains quantized even in situations, where the $\theta$-term in the bulk and the associated surface Hall conductivities, $\sigma_{xy}^S$, are not quantized due to the breaking of time-reversal symmetry.
The quantization arises as the $\theta$-term changes by $\pm 2 \pi n$ along a loop around $n$ edge channels.
Model calculations show how an interplay of orbital and Zeeman effects leads to quantum Hall transitions, where channels get redistributed along the edges of the crystal.
The network of edges opens new possibilities to investigate the coupling of edge channels.
\end{abstract}


\maketitle


In topological insulators (TIs) the topological properties of the band structure leads to the formation of protected metallic states at the surface.
Soon after their theoretical prediction\cite{Kane2005, Bernevig2006} as a generalization of Haldane's Chern insulator\cite{Haldane1988} their two-dimensional (2d) variant --- the quantum spin Hall insulator --- was realized in quantum well heterostructures by K\"onig \textit{et al.}\cite{Koenig2007}.
Later, three-dimensional (3d) varieties of this novel form of insulator were predicted\cite{Fu2007, Moore2007, Roy2009} and again realized afterwards\cite{Hasan2009, Bruene2011}.
3d TIs come as strong and weak TIs.
Strong TIs (STIs) have a metallic surface which is protected against localization due to its helical nature.
Band topology enforces that the surface metal can be described as an effective 2d Dirac theory of massless electrons.
Soon after the realization of 3d STIs it was understood that the surfaces of these materials could give way to an unconventional Hall response whenever the surface is gapped by perturbations which break time-reversal symmetry (TRS).
This is related to the axion quantum electrodynamics\cite{Wilczek1987, Qi2008, Essin2009} and the variation of the $\theta$-angle at an interface between a bulk material ($\theta \neq 0$) and vacuum ($\theta = 0$).
STIs are characterized by the quantized value $\theta = \pi$.
When a gap is opened, a dissipationless surface conductivity with a half-integer Hall conductivity\cite{Qi2008, Essin2009}, $\sigma^{S}_{xy}=\frac{e^2}{2h}$, arises.
Importantly, this value is \emph{not}\ quantized if both TRS and inversion symmetry (IS) are broken in the bulk.

Recently, Br\"une \textit{et al.}\cite{Bruene2011} have experimentally observed a quantized Hall effect for films of HgTe which can be considered 3d STIs, with the magnetic field applied perpendicular to the film.
Both in the experiment and in two theoretical studies\cite{Lee2009, Chu2011, Zhang2011} a field configuration was used, where only two faces of the crystal were gapped.

In this paper we investigate the Hall response of a 3d STI in the presence of a bulk magnetic field oriented such that all surface excitations are gapped.
We first study how the $\theta$-term characterizing the 3d bulk, the surface Hall conductivities, the edge channels, and the quantum Hall effect observed by contacting those edge channels are related to each other.
We then study a concrete lattice model (inspired by the HgTe band structure) to investigate how the location and number of edge channels can be controlled.


We consider a single crystal STI with planar surfaces in the presence of a magnetic field in a configuration, where both the bulk and all surfaces --- up to one-dimensional edge channels discussed below --- are gapped.
The macroscopic electrodynamic response is described by an effective topological field theory in (3+1) dimensions defined by the action\cite{Qi2008, Essin2009, Essin2010, Qi2010}
\begin{align}
\label{eq:AxionED_1}
S_{\theta} &= \frac{e^2}{4 \pi h c} \int d^3r \, dt \, \theta \, \epsilon^{\mu \nu \sigma \tau} \partial_\mu A_\nu \partial_\sigma A_\tau \\
\label{eq:AxionED_2} \nonumber
&= \frac{e^2}{2 \pi h c^2} \int d^3r \, dt \, \theta \, \vect{E} \cdot \vect{B}
\end{align}
This term is referred to as the $\theta$-term, well-known from the field theory of axion electrodynamics\cite{Wilczek1987}.
Several properties\cite{Qi2008, Essin2009, Essin2010} are important in the following:
\textit{(i)}\ The value of $\theta$ is defined modulo $2 \pi$.
\textit{(ii)}\ $S_{\theta}$ is an integral over a total derivative, \textit{i.e.}, it has no effect for $\theta = \mathrm{const.}$, but matters at interfaces and surfaces, where $\theta$ changes (see below).
\textit{(iii)}\ Within the bulk, $\theta = \theta_{\mathrm{b}}$ is constant (we define $0 \le \theta_{\mathrm{b}} < 2 \pi$ to avoid ambiguities).
In the presence of either IS or TRS, $\theta_{\mathrm{b}}$ assumes the values $0$ or $\pi$.
As TRS band insulators have $\theta_{\mathrm{b}} = 0$ and STIs have $\theta_{\mathrm{b}} = \pi$, this can be used to classify STIs (even in the presence of interactions).
In the absence of both IS and TRS, $\theta_{\mathrm{b}}$ is not quantized, and a TI can adiabatically be transformed into a band insulator.

Varying Eq.~(\ref{eq:AxionED_1}) with respect to $A_\mu$ one obtains dissipationless currents
\begin{equation}
\label{eq:Current}
\vect{j} = -\frac{e^2}{2 \pi h} \vect{\nabla} \theta(\vect{r}) \times \vect{E}
\end{equation}
located at the surface across which $\theta$ changes.
For a homogeneous electric field at a surface $S_i$ one obtains the surface Hall conductivity by integrating the current perpendicular to $S_i$.
Here it is important to realize that in the vacuum $\theta = 2 \pi n$ with integer $n$.
We find it convenient to associate with each surface $S_i$ an integer $n_i$ defined by the value of $\theta$ obtained by approaching the surface (see below for a numerical calculation).
Defining the Hall conductivity $\sigma^H_i$ of $S_i$ by measuring the current in the direction $\hat{\vect{n}}_i \times \vect{E}$ for an electric field parallel to the surface, one obtains from Eq.~(\ref{eq:Current}):
\begin{equation}
\label{eq:SurfaceHallConductivity}
\sigma^H_i = \frac{e^2}{h} \left( n_i - \frac{\theta_{\mathrm{b}}}{2 \pi} \right).
\end{equation}
For $\theta_{\mathrm{b}} = 0$ one recovers the integer quantization of a conventional quantum Hall system and for $\theta_{\mathrm{b}} = \pi$ the well known shift obtained for the 2d Dirac equation.
In general, any value of $\theta_{\mathrm{b}}$ is allowed.
For a 3d sample with $N$ surfaces the $N$ integers $n_i$ and the constant $\theta_{\mathrm{b}}$ completely characterize the macroscopic Hall response.

As a simple application we can consider a quasi-2d slab with surfaces $S_1$ and $S_2$, and an electric field which is the same on the top and bottom of the sample.
As $\hat{\vect{n}}_1 = -\hat{\vect{n}}_2$, the total 2d Hall response is given by
\begin{equation}
\label{eq:Sigma2D}
\sigma_H^{\mathrm{2d}} = \sigma^H_{1} - \sigma^H_{2} = \frac{e^2}{h} (n_1 - n_2),
\end{equation}
yielding the well-known integer-quantized form which is independent of the non-quantized $\theta_{\mathrm{b}}$.

\begin{figure}[tb]
\includegraphics[width=0.9 \linewidth]{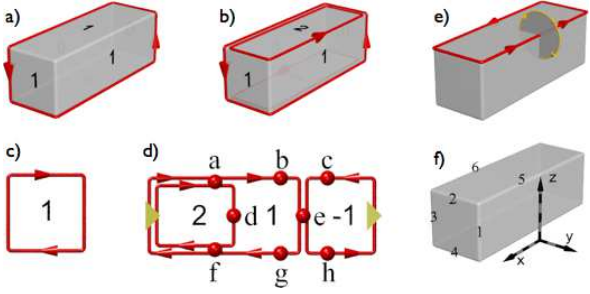}
\caption{%
(Color online)
a,b) Two possible edge channel configurations which are projected onto a plane in c) and d), respectively.
On each face the integer $n_i$ is written which characterizes the surface and from which the number of edge channels can be obtained using Eq.~(\ref{eq:WindingNumber}) (for the planar graphs, $n_i = 0$ on the outside).
The two configurations can be obtained by small rotations of the field, see Fig.~\ref{fig4}.
e) Orientation of line integrals used in Eq.~(\ref{eq:WindingNumber}).
f) Numbering of edges and coordinate system.
}
\label{fig1}
\end{figure}

This analysis does, however, not answer the question what is actually measured when metallic contacts are attached to a 3d sample to obtain the Hall response.
Interestingly, Chu, Shi and Shen\cite{Chu2011} argue that in the presence of \emph{dissipative}\ surfaces (not considered by us) it might be possible to observe directly an approximately half-quantized $\sigma^H_i$ in a four-terminal measurement.
But in general, neither $\sigma^H_i$ nor $\sigma_H^{\mathrm{2d}}$ are measured.
Instead, one has to investigate a network of gapless current-carrying 1d channels (also bent, non-planar quantum Hall junctions show similar networks\cite{Grayson2007}).
Each chiral 1d channel carries exactly one conductance quantum, $I = \frac{e^2}{h} \Delta \Phi$, where $\Phi$ is the scalar potential.
To find those, we use Eq.~(\ref{eq:Current}) to calculate the current through a 2d manifold whose boundary is given by an equipotential line, $\Phi(\vect{r}) = \Phi_{\mathrm{e}}$.
From Stokes' theorem one obtains assuming that $\theta(\vect{r})$ is singular at most at one point on the manifold
\begin{equation}
\label{eq:WindingNumber}
I = \frac{e^2}{h} \mathcal{W} \Delta \Phi, \quad
\mathcal{W} = \frac{1}{2 \pi} \oint d\vect{r} \cdot \vect{\nabla} \theta = n_i - n_j
\end{equation}
with $\Delta \Phi = \Phi_{\mathrm{s}} - \Phi_{\mathrm{e}}$, where $\Phi_{\mathrm{s}}$ is the potential at the singularity.
Therefore, the number of 1d channels is given by the (integer) winding number $\mathcal{W}$ of the $\theta$-term, and their location tracks the singularities in $\theta(\vect{r})$ associated with $\mathcal{W} \neq 0$.
The line integral is taken along the boundary of the manifold in clockwise direction looking along the direction of $I$ [see Fig.~\ref{fig1}~e)].
For crystals with flat surfaces, the channels are localized along the edges, and $\mathcal{W}$ is obtained from the integers characterizing the adjacent surfaces, $\mathcal{W} = n_i - n_j$, where $i$ ($j$) is the surface with surface normal parallel (antiparallel) to the integration direction, respectively.

Note that the winding numbers are quantized for \emph{any}\ value of $\theta_{\mathrm{b}}$.
This reflects that the quantization of the Hall effect (based on charge quantization) is a more robust concept than the $\mathbb{Z}_2$ quantization used to classify TIs (which relies on TRS).
Deforming the TI into a ring (with a macroscopic cross section) one can directly apply Laughlin's gauge argument\cite{Laughlin1981} for an alternative proof of the quantization of the Hall effect measured by contacting those edge channels.

To determine how different networks of edge channels can be realized, we study a concrete example.
As a minimal model\cite{Fu2007} for a TI we consider four bands in a cubic lattice using orbitals inspired by strained 3d HgTe\cite{Bruene2011} (different orbitals are important for 2d HgTe).
We introduce four basis states\cite{Dai2008} within the unit cell:
\begin{equation*}
\begin{aligned}
|1\rangle &= |E1, \uparrow \rangle = |s, \uparrow \rangle, \quad
|3\rangle  = |E1, \downarrow \rangle = |s, \downarrow \rangle, \\
|2\rangle &= |LH, \uparrow \rangle = (|p_x + i p_y, \downarrow \rangle - 2 |z, \uparrow \rangle) / \sqrt{6}, \\
|4\rangle &= |LH, \downarrow \rangle = -(|p_x - i p_y, \uparrow \rangle + 2 |z, \downarrow \rangle) / \sqrt{6}.
\end{aligned}
\end{equation*}
In this basis, TRS is implemented by $\hat{\Theta} = -i \sigma_z \otimes s_0 \hat{K}$, where $\sigma_z$ acts in spin space, and the $2 \times 2$ identity matrix $s_0$ acts on the orbital degrees of freedom ($\hat{K}$ is complex conjugation, $\sigma$ and $s$ are standard Pauli matrices).
Parity is implemented via $\hat{P} = \sigma_0 \otimes s_z$.
Two-fold rotation symmetries along the $x$-, $y$-, and $z$-directions are generated by $\hat{R}_x(\pi) = i \sigma_x \otimes s_z$, $\hat{R}_y(\pi) = -i \sigma_y \otimes s_z$, and $\hat{R}_z(\pi) = i \sigma_z \otimes s_0$.
The $4 \times 4$ Hamiltonian can conveniently be expressed in terms of the identity, $\Gamma^0$, five Dirac matrices $\Gamma^a$, and their ten commutators $\Gamma^{ab} = [\Gamma^a, \Gamma^b]/(2i)$ which satisfy the Clifford algebra, $\{ \Gamma^a, \Gamma^b \} = 2 \delta_{a,b} \Gamma^0$.
Using $\Gamma^{(1,2,3,4,5)} = (\openone \otimes s_z, -\sigma_y \otimes s_x, \sigma_x \otimes s_x, -\openone \otimes s_y, \sigma_z \otimes s_x)$ we find as a minimal model
\begin{align}
\label{eq:EffectiveHam}
H =& -t \sum_{n, j=1,2,3} \left( \Psi^\dagger_{n+\hat e_j} \frac{\Gamma^{1} - i \Gamma^{j+1}}{2} e^{i A_{n,n+\hat e_j}} \Psi^{\phantom{\dagger}}_{n}  + \mathrm{H.c.} \right) \nonumber \\
& +  \sum_n \Psi^\dagger_n  ( m \Gamma^{1}+\Delta_1 \Gamma^5 + \Delta_2 \Gamma^{15})\Psi^{\phantom{\dagger}}_n + H_Z
\end{align}
where $m$ is the tuning parameter and $t=1$ is the hopping amplitude.
$\Delta_1$ and $\Delta_2$ parametrize the breaking of IS, $\Delta_1$ also breaks TRS.
For systems without inversion symmetry (as HgTe), $\Delta_2 \neq 0$ and a finite $\Delta_1$ will be induced by magnetic fields.
Orbital effects of the magnetic field are described by the Peierls factor $\exp(i A_{n, n + \hat{\vect{e}}_j})$ defined on the link from site $n$ to the neighboring site $n + \hat{\vect{e}}_j$.
The Zeeman effect is obtained from
\begin{multline}
H_Z = \sum_n \Psi^\dagger_n \bigl[ -B_Z^x (g^+ \Gamma^{25} + g^- \Gamma^{34}) \\
- B_Z^y (g^+ \Gamma^{35} - g^- \Gamma^{24}) + B_Z^z (g^+ \Gamma^{23} + g^- \Gamma^{45}) \bigr] \Psi^{\phantom{\dagger}}_{n},
\end{multline}
where $\vect{B}_Z = \mu_B \vect{B}$ and $g^\pm = (g_{E1} \pm g_{LH})/2$ are linear combinations of the $g$ factors of the $E1$ and $LH$ subbands.
For our plots we use $m = -2$, $\Delta_2 = 0$ ($\Delta_2 \neq 0$ does not change our results qualitatively), $g^{+} = g^{-} = 1$, and for a magnetic field in the $yz$-plane we use the Landau gauge, \textit{i.e.}, $A_{n, n + \hat{\vect{e}}_j} = 2 \pi (\Phi_y z_n - \Phi_z y_n) \delta_{\hat{\vect{e}}_j, \hat{\vect{e}}_x}$, where $\Phi_{i}$ is the flux per unit cell in direction $i$ in units of the flux quantum.

\begin{figure}[tb]
\includegraphics[width=0.99 \linewidth]{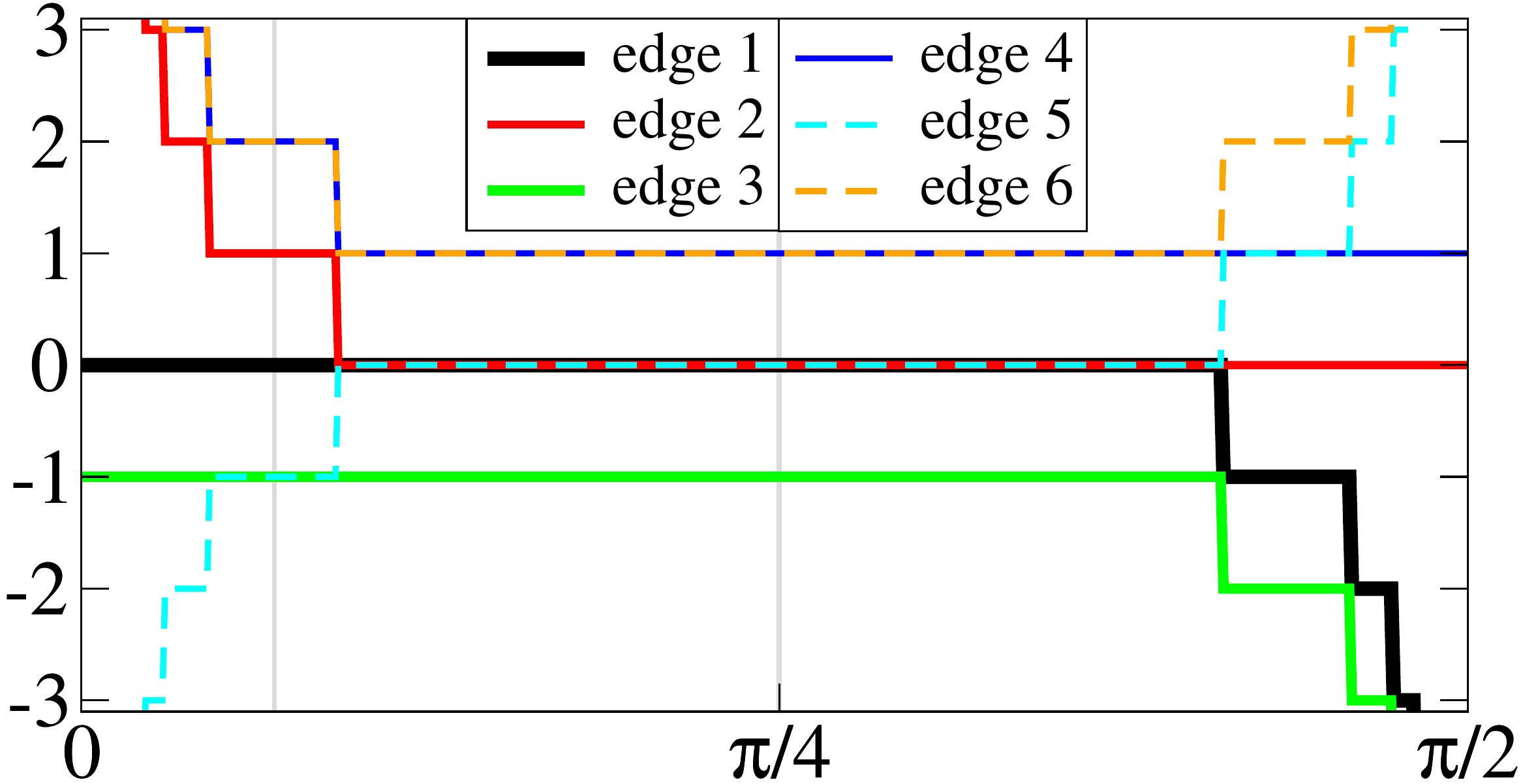}
\caption{%
(Color online)
Number of edge channels along edges 1--6 defined in Fig.~\ref{fig1}~e) for $\mu = 0.02$, $\Delta_1=0$, $B = 0.001 (1, \cos \varphi, \sin \varphi) / \sqrt{2}$, as a function of the rotation angle $\varphi$ obtained from Eqs.~(\ref{eq:ZeemanMass}, \ref{eq:EnergyLevels}, \ref{eq:EdgeChannelNumbers}).
The sign indicates whether the current runs parallel or antiparallel to the $\hat{\vect{x}}, \hat{\vect{y}}$ and $\hat{\vect{z}}$ directions.
For the two configuration indicated by gray vertical lines, the edge configuration is shown in Fig.~\ref{fig1}~a) and b).
The number of channels on the six other edges is obtained by inversion symmetry (assumed here).
}
\label{fig2}
\end{figure}

For $\Delta_1 = \Delta_2 = 0$ and vanishing magnetic field, the model describes a trivial band insulator for $|m| > 3$ and a STI for $0 < |m| < 3$.
At $m = \pm 3$ the bulk band structure is characterized by a 3d Dirac point.
For $0 < \Delta m = m + 3 \ll 1$ and $\Delta_i, B_Z^i, \Phi_i \ll \Delta m$ one can obtain analytically the theory for a surface with normal vector $\hat n$.
First, one obtains the surface state for vanishing $\Delta_i, B_Z^i, \Phi_i$ from the standard bound state solution of the 3d Dirac equation\cite{Hasan2010} with space dependent mass $m_D = -\Delta m \, \sign(\hat{\vect{n}} \cdot \vect{r})$, where a negative $m_D$ for $\hat{\vect{n}} \cdot \vect{r} > 0$ mimics the vacuum.
From this one obtains the 2d Dirac equation describing the surface of a TI\cite{Hasan2010}.
To linear order, the effects of $\Delta_i$ and $B_Z^i$ are just given by the projection of $H$ on the surface state, while the orbital effects are covered by minimal substitution.
We obtain
\begin{eqnarray}
\label{eq:SurfaceTheory}
H^{\mathrm{surf}} &= v_F \vect{\sigma} \cdot \left( \vect{p} - \frac{e}{c} \vect{A} \right) - \sigma_0 \mu + m_Z \sigma^3 \\
\label{eq:ZeemanMass}
m_Z &= \hat n \cdot(-g^- B_Z^x,- g^- B_Z^y, g^+ B_Z^z) - \Delta_1
\end{eqnarray}
where $\vect{\sigma} = (\sigma^1, \sigma^2)$, $\vect{p} = (p_x, p_y)$ denote the momenta perpendicular to $\hat{\vect{n}}$ (shifted by a $\vect{B}_Z$ dependent constant), $\sigma^3$ is the Pauli matrix parallel to $\hat{\vect{n}}$, and $\mu$ denotes the chemical potential.
$\vect{A}$ is the vector potential of the normal field component appropriately transferred into the new basis.
Note the unexpected dependence of $m_Z$ on the direction of the magnetic field which allows to control orbital and Zeeman effects independently.
For example, $m_Z$ can be finite even for a magnetic field perpendicular to the surface.

It is an elementary exercise\cite{Haldane1988} to compute the spectrum of the massive Dirac theory in a field~(\ref{eq:SurfaceTheory}):
\begin{equation}
\label{eq:EnergyLevels}
E_j = \sign(j)  \sqrt{\omega_c^2 |j| + m_Z^2},
\; E_{j=0} = \sign(B_\perp) m_Z.
\end{equation}
with integer $j$ and $\omega_c = v_F \sqrt{2 |e B_{\perp}|}$.
$B_\perp = \vect{B} \cdot \hat{\vect{n}}$ determines the orbital contribution of the magnetic field.
As long as $\vect{B}$ is not almost parallel to the surface, $m_Z$ is only important for $j = 0$ as $m_Z^2 \sim (\mu_B B)^2 \ll v_F^2 |e B_{\perp}|$.

As the levels~(\ref{eq:EnergyLevels}) do not cross, the results for the integers $n_i$ characterizing the surfaces, do not depend directly on $m_Z$ and are obtained from the QHE for Dirac fermions\cite{Haldane1988} and Eq.~(\ref{eq:SurfaceHallConductivity}):
\begin{equation}
\label{eq:EdgeChannelNumbers}
n = \begin{cases}
j+1, & B_\perp > 0 \\
-j,  & B_\perp < 0
\end{cases} \quad \text{for $E_j < \mu < E_{j+1}$}.
\end{equation}
Using Eqs.~(\ref{eq:ZeemanMass},\ref{eq:EnergyLevels}, \ref{eq:EdgeChannelNumbers}) one can therefore calculate the relevant winding numbers which determine the position of the edge channels, see Eq.~(\ref{eq:WindingNumber}) and Fig.~\ref{fig1}.
Note that $\mu$ can be controlled independently on each surface by appropriate gates and might also depend on details of the surface chemistry.
As an example, we show in Fig.~\ref{fig2} how the number of edge channels changes when a magnetic field is rotated around the $x$ axis.
This allows to realize a wide variety of channel configurations.

\begin{figure}[tb]
\includegraphics[width=0.99 \linewidth]{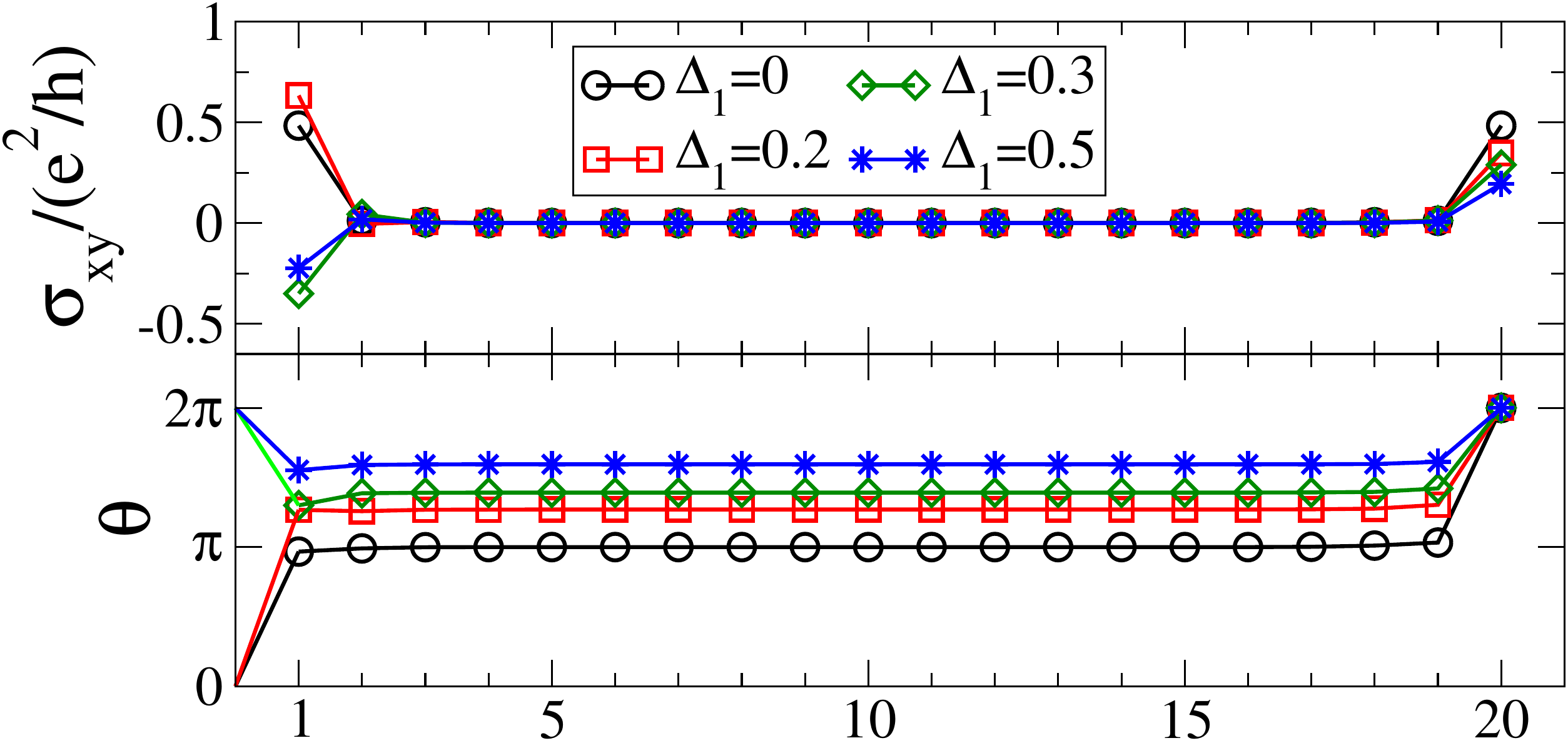}
\caption{%
(Color online)
$\sigma_{xy}(i)$ and $\theta_i = 2 \pi j + \frac{\hbar}{e^2} \sum_{j \le i} \sigma_{xy}(j)$ as a function of distance from the surface for an infinite slab in the $xy$ plane (20 layers, $\Phi_i = 0$, $g^+ B_Z = 0.25$).
The integer $j$ is chosen such that $0 \le \theta <2 \pi$ in the center.
While each surface conductivity $\sigma^H_i$ and $\theta_{\mathrm{b}}$ are not quantized, $\sigma_H^{2d} = \sum_i \sigma_{xy}(i)$ and, equivalently, $\mathcal{W} = \theta(20) - \theta(0)$ are quantized.
$\mathcal{W}$ jumps from $2 \pi$ to $0$ at $\Delta_1 \approx g^+ B_Z^z = 0.25$, where $m_Z$ in Eq.~(\ref{eq:ZeemanMass}) changes sign and one surface undergoes a QH transition.
}
\label{fig3}
\end{figure}

The low-energy theory~(\ref{eq:SurfaceTheory}) with infinite cutoff does not account for the change of the $\theta$-term in the bulk, while it can be computed numerically by considering the full model~(\ref{eq:EffectiveHam}).
For an infinite slab in the $xy$-plane, we have calculated from the Kubo formula the matrix of conductivities, $\sigma_{xy}(i,j)$, describing the current in the $x$ direction in layer $i$ when an electric field is applied in layer $j$ in the $y$ direction.
For simplicity, we considered here only a Zeeman magnetic field.
$\sigma_{xy}(i,j)$ is to a good approximation diagonal (as implicitly assumed in Eq.~\ref{eq:AxionED_1}), and we plot in Fig.~\ref{fig3} $\sigma_{xy}(i) = \sum_j \sigma_{xy}(i,j)$, \textit{i.e.}, the current distribution for a uniform electric field, and also the corresponding layer-resolved $\theta_i$ obtained from Eq.~(\ref{eq:Current}).
For finite $\Delta_1$ such that TRS and IS are broken, $\theta_{\mathrm{b}}$ and the surface conductivities are not quantized, but the total 2d Hall response remains quantized as in Eq.~(\ref{eq:Sigma2D}).

\begin{figure}[tb]
\begin{center}
\includegraphics[width=0.47\textwidth]{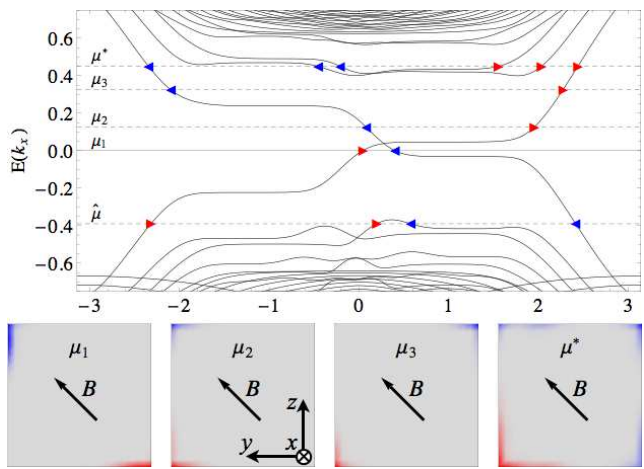}
\end{center}
\caption{%
(Color online)
Band structure of a beam of dimensions ($\infty \times 20 \times 20$) as function of the momentum $k_x$ along the beam ($\Phi_y = \Phi_z = 0.018$, $g^- B_Z^y = g^+ B_Z^z = 0.28$,  $\Delta_1 = 0.1$).
Flat parts in the central region represent Landau levels while dispersive parts are one-dimensional edge channels.
Lower panel: Position of the edge channels for four chemical potentials marked above.
Red (blue) denotes edge channels running parallel (antiparallel) to the $\hat{\vect{x}}$ direction.
For $\mu \approx 0$, when $m_Z$ dominates, $\vect{B}$ points towards/away from the edge channels, while a more conventional perpendicular arrangement is obtained in the orbitally dominated regime, $\mu_3$.
The last figure shows three edge modes moving in $\hat{\vect{x}}$ direction on one edge, while three $-\hat{\vect{x}}$ modes are distributed over the other three edges.
Also two counterpropagating channels at the same edge can be obtained, e.g., for $\hat{\mu}$ (not shown).
}
\label{fig4}
\end{figure}

To illustrate the localization of the edge channels along the edges of the 3d crystal and to show how our findings relate to the band structure, we have calculated the properties of an infinite beam of dimension $\infty \times 20 \times 20$ using exact diagonalization.
In the appropriate Landau gauge, the momentum along the beam is conserved and one obtains the band structure of Fig.~\ref{fig4}~a).
To obtain the position of the conducting channels we plot in Fig.~\ref{fig4} the corresponding probability distribution: as expected they are located at the edges of the sample.
By changing the chemical potential a sequence of quantum Hall transitions is obtained.
The first three figures show how the edge channels jump from one edge to the other, conserving the number of channels.
For other cases, the number of channels can also be increased as in a conventional quantum Hall transition and one can obtain more complex situations, see last panel.


For a given edge channel configuration one can determine the conductance measured when contacting those channels using (i) Eq.~(\ref{eq:WindingNumber}), (ii) Kirchhoff's law (current conservation at each corner) and (iii) that the electrochemical potential remains constant when channels split.
For the circuit shown in Fig.~\ref{fig1}~d) one obtains $\Phi_a = \Phi_b = \Phi_d$, $\Phi_e = \Phi_g = \Phi_h = \Phi_a - \frac{h}{e^2} I$, $\Phi_c = \Phi_a - 2 \frac{h}{e^2} I$, and $\Phi_f = \Phi_a - \frac{h}{2 e^2} I$.
Defining the cross-conductances $G_{\alpha \beta} = I/(\Phi_{\beta} - \Phi_{\alpha})$, this implies that not only, e.g., $G_{gb} = e^2/h$ and $G_{fa} = 2 e^2/h$ can be measured but also $G_{cf} = \frac{2}{3} e^2/h$.
At the points where two channels meet, the power $P_i$ is dissipated, with a quantized ohmic conductance $I^2/P_i$ which takes the values $2 e^2/h$ and $\frac{1}{2} e^2/h$ at the `merging' corners above contacts e and f, respectively.
Such a quantized heat source might have interesting applications for the study of heat transport in TIs.

Can a topological insulator uniquely be identified by measuring the quantum Hall effect?
For a fixed magnetic field this is not possible since there are situations where, for example, all surfaces can be gapped and there is not a single edge channel anywhere.
The changing pattern of edge channels when field strength or direction or gate voltages are varied, can, however, serve as a fingerprint of topological insulators and their Dirac metal.
The network of edge channels also opens new experimental opportunities to study quantum Hall physics by separately addressing edge channels.
Another interesting problem is the physics of fractional quantum Hall states, especially in combination with a non-trivial topology of surfaces and edges obtained by drilling holes into a 3d topological insulator.

We acknowledge discussions with A.~M.~Essin,  E.~M.~Hankiewicz, M.~Garst, M.~A.~Levin, N.~Nagaosa, and M.~Vojta.

Our support came from DFG's FOR 960, SFB 608, and the Bonn-Cologne Graduate School (MS).

\end{document}